\documentclass[aps,pre,twocolumn,floatfix]{revtex4}
\usepackage{graphicx}

\begin{document}

%\setcounter{chapter}{1}

%\chapter{UNDER-KNOTTED AND OVER-KNOTTED POLYMERS: COMPACT SELF-AVOIDING LOOPS}
\title{Under-knotted and over-knotted polymers: compact self-avoiding loops}

%\markboth{R.C. Lua, N.T. Moore, A.Y. Grosberg}{Under-Knotted and
%Over-Knotted Polymers:  2. Compact self-avoiding loops}

\author{Rhonald C. Lua, Nathan T. Moore, Alexander Yu. Grosberg}

\affiliation{Department of Physics, University of Minnesota\\
Minneapolis, MN 55455, USA\\
E-mail: grosberg@physics.umn.edu}

\begin{abstract}

We present a computer simulation study of the compact self-avoiding
loops as regards their length and topological state.  We use
a Hamiltonian closed path on the cubic-shaped segment of a $3D$ cubic
lattice as a model of a compact polymer.  The importance of ergodic
sampling of all loops is emphasized.  We first look at the effect
of global topological constraint on the local fractal geometry of
a typical loop.  We find that even short pieces of a compact
trivial knot, or some other under-knotted loop, are somewhat crumpled
compared to topology-blind average over all loops.  We further
attempt to examine whether knots are localized or de-localized
along the chain when chain is compact.  For this, we perform
computational decimation and chain-coarsening, and look at the
"renormalization trajectories" in the space of knots frequencies.
Although not completely conclusive, our results are not inconsistent
with the idea that knots become de-localized when the polymer is
compact.

\end{abstract}
\maketitle
\section{Introduction}     %S1-Heads

\subsection{Goal and plan of this work}

In the preceding article\cite{preceding}, we have discussed the
idea of two distinct scaling regimes for the closed polymer loop,
we called them the over-knotted and under-knotted regimes.  They
are different in terms of the relation between a given loop with
quenched topology and the imaginary loop with annealed topology,
the latter being able to freely cross itself.  Loosely speaking,
for the over-knotted loop, topological constraints prevent the
loop from untying its knots, from simplifying the knots or
reducing their numbers.  For the under-knotted loops, the effect
is the opposite, for the topological constraints prevent the loop
in this case from forming more knots.  We argued that
under-knotted loops tend to swell beyond the size of the phantom
loop, while over-knotted loops tend to be more compact.

In the present article, we wish to extend this qualitative idea
for another context, namely, for the polymers which are compressed
by either external pressure or by some sort of poor solvent
effect, the latter being equivalent to somewhat sticky segments.
This problem is highly relevant, because most biopolymers are
compact in their native conditions.  DNA, for instance, is
frequently so long that it would not fit inside the cell if
allowed to fluctuate freely.

We shall argue first of all that the natural zeroth approximation
for such a problem of compact loops can be formulated in terms of
Hamiltonian paths on cubic shaped piece of a regular lattice in
space.  In the following sections we shall briefly describe our
approach to computational generation of the maximally compact
loops on a cubic lattice.  This approach was formulated in the
work\cite{1}.  Next we present probabilities of obtaining
unknotted configurations, as well as probabilities of obtaining a
few simple knots. We also present measurements of the average
spatial extent of segments or sub-chains of compact conformations
and analyze how these measurements depend on the topology or
knot-type of the conformations.

Through the latter measurements, we address the question how
global knot topology of a large loop affects local fractal
geometry of typical trajectories.  An exciting general problem
here, which we think is an important challenge for the future, is
to which extent looking at local geometry of chains allows
guessing of their global topology.

We should point out that compact loops very rapidly become
under-knotted with increasing loop length with fixed knot state,
because the ensemble of all loops (or, equivalently, the ensemble
of states of a phantom loop) is dominated by very complex knots
when loop is compact.  Thus, what we study is the local fractal
geometry of under-knotted loops.  This already suggests the
result, which we shall explain in more depth later, that the
pieces of under-knotted compact loop are locally somewhat crumpled
compared to their phantom counterparts.

There is also an interesting question of whether the knots tend to
be spread out or confined to a small portion of the polymer.
Previous theoretical and computational work\cite{knot_inflation,6,localization_2,localization_3}
have shown that knot-determining domains for non-compact loops are
usually rather tight. For instance, the preferred size of the
trefoil-determining portions of knotted polymer chains corresponds
to just seven freely jointed segments\cite{6}. In that work, the
knotted domain is identified by looking for the minimal number of
contiguous segments belonging to the circular chain such that,
upon closure with an external planar loop, the new knot formed is
of the same type as the original knot.

One possible interpretation would be that knot localization is the
property of all under-knotted loops.  From that point of view, it
is interesting to look at knot localization in compact loops, as
they are heavily under-knotted.  So far, we are aware of only one
numerical work\cite{7} addressing this issue for prime \emph{flat}
knots in a model of self-attracting polymers with excluded volume.
Here, when we say "flat" we mean that the polymer is strongly
adsorbed onto a flat surface, or confined in a thin planar slit.
It was found in\cite{7} that these flat knots are localized in the
high temperature swollen regime (consistent with theoretical
prediction\cite{localization_3}), but become delocalized in the
low temperature, collapsed globular phase.  For three-dimensional
polymers, the conjecture is that there is also a similar
delocalization transition, i.e. the knots are delocalized for
compact circular chains.  We shall attempt to address this issue
of knot (de)localization for compact circular chains.

\subsection{Why lattice model is natural for our purposes}

In order to address knots in densely compressed loops it is very
important to realize the role of excluded volume, or
self-avoidance.  Physically, this is the short range repulsive
forces which always exist and which prevent pieces of polymer from
penetrating each other.  Mathematically, self-avoidance condition
specifies the ensemble of allowed loop shapes.  When loop is not
geometrically restricted, as it was examined in the preceding
article\cite{preceding}, these excluded volume constraints are
often irrelevant.  That is why, for the purposes of the preceding
paper\cite{preceding}, we argued it possible to view polymer loop
as just a closed continuous mathematical line, with no thickness.
For such object, the measure of trajectories with self-overlaps is
exactly zero, and so the probabilities of all distinct knot
topologies sum up to unity.  For the compact loop, such model
becomes meaningless, because the limit of zero thickness is
singular for loops restricted to within certain volume.  The
meaningful model in this case implies that the density is
maintained constant all across the allowed volume, which exactly
corresponds to imposing the self-avoidance condition along with
the volume restriction.

The simple model capturing self-avoidance condition is a polymer
presented as a path on the regular lattice in space, such as cubic
lattice.  Knots in lattice polymers were examined in a number of
works.  In particular, it was proven in the
works\cite{theorem1,theorem2} that the probability to obtain a
trivial knot upon random generation of the lattice polygon of $N$
segments decays exponentially in $N$.  We should emphasize that
this is different from the problem addressed in our preceding
article\cite{preceding}, because we looked at the loops with no
volume exclusion, while every lattice model suitable for the study
of knots and thus excluding self-intersections involves
automatically some excluded volume.  Unlike geometrically
unrestricted loops, for the problem of compact loops excluded
volume is what we must take into account, and, therefore, lattice
path is now a natural model to look at.

In order to make it compact, we now consider a segment of cubic
lattice, say, a cube of the some size $m \times m \times m$, and
consider a path of $N = m^3$ self-avoiding steps confined in this
cube.  Obviously, this is a Hamiltonian path.  For the purposes of
modelling the closed polymers, we shall consider here Hamiltonian
closed paths, or Hamiltonian loops.

\section{Brief overview of our recent results\protect\cite{1}}

\begin{figure}[th]		%Fig~8
%\centerline{\psfig{file=AGLatticeLoop.eps,width=3.6in,angle=0}}
\centerline{\scalebox{0.3}{\includegraphics{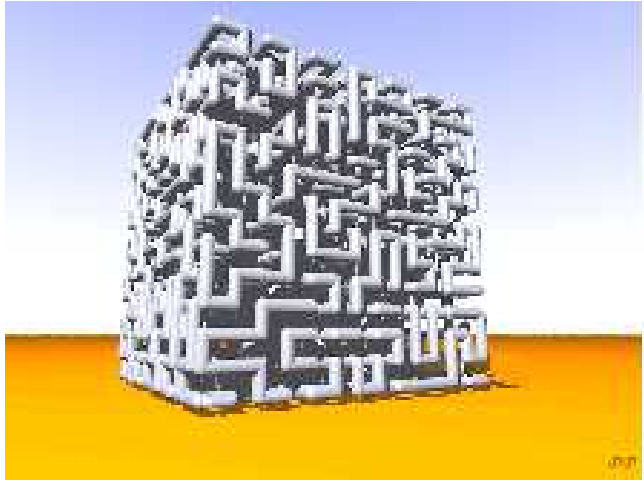}}}
\vspace*{8pt}
\caption{A $14\times14\times14$ compact conformation on a lattice rendered by Professor
Piotr Pieranski. The three knot invariants calculated for this conformation correspond to a trivial knot.
Using a knot-tightening algorithm called Shrink On No Overlaps\protect\cite{SONO} (SONO),
Professor Pieranski verified that the conformation is indeed unknotted by reducing
the conformation into a small circular loop.}\label{fig8}
\end{figure}

\subsection{Generation of compact loops}
The method we used to generate compact conformations on a lattice
is described in Ref.~\cite{1}. It is based on a combinatorial
algorithm by Ramakrishnan {\it et al}\cite{2}. The method
essentially works by placing links randomly on the lattice,
avoiding sub-cycles and dead ends, until a single loop fills the
desired cubic lattice dimensions. For closed loops (Hamiltonian
walks), conformation dimensions can only be even ($2 \times 2
\times 2$, $4 \times 4 \times 4$, etc.). Note that the method does
not use a conventional process in which a single connected chain
is grown to the desired size, because the rejection rate of any
such process for compact chains rapidly becomes catastrophic even
at modest chain length. Figure~\ref{fig8} illustrates an example of a $14\times14\times14$ conformation.
Although our method is not free of biases
and is not perfect, it is a significant improvement over the
original algorithm by Ramakrishnan {\it et al}.

\subsection{Topology}
We identified the knot-type (${\cal K}$) of a conformation by
calculating the following three knot invariants: the Alexander
polynomial evaluated at $t=-1$ ($\Delta(-1)_{\cal K}$), the
Vassiliev invariant of degree two ($v_2({\cal K})$) and the
Vassiliev invariant of degree three ($v_3({\cal K})$). For
example, the knot invariants for an unknot or trivial knot (${\cal
K}\equiv0_1$) are $|\Delta(-1)_{0_1}|=1,v_2(0_1)=0,v_3(0_1)=0$.
Although, it is possible for two distinct knots to have the same
set of knot invariants, we expect the false identification of
knots to be rare. For instance, the set of three knot invariants
are distinct from those of (prime) knots with 10 crossings or
fewer (249 knots in all) in their projection.

\begin{figure}[th]      %Fig~1
%\centerline{\psfig{file=knotprobfinal.eps,width=2.6in,angle=0}}
\centerline{\scalebox{1}{\includegraphics{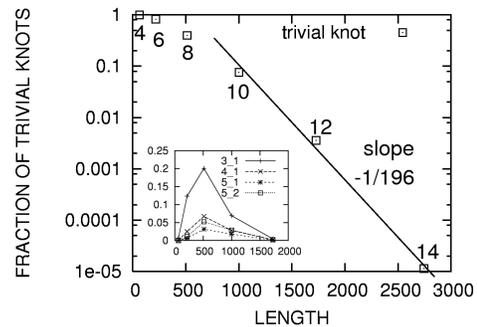}}}
\vspace*{8pt}
\caption{Trivial knot probabilities for compact conformations of size $4\times4\times4$ to $14\times14\times14$.
Inset shows the probabilities of the non-trivial knots $3_1$ (trefoil), $4_1$ (figure-eight), $5_1$ (star),
$5_2$.}\label{fig1}
\end{figure}

Using these knot invariants to classify the conformations, we
collected data for the frequency of occurrence of the trivial knot
and the first few simple knots (Figure~\ref{fig1}). Computational
data on trivial knot probability are customarily fit to
exponential, our last three data points giving $\sim
\exp(-N/196)$.

An exponential fit should not be surprising, as the total number
of conformations of length $N$ grows exponentially with $N$.  The
estimate of total number of compact conformations can be read out
of the Flory\cite{floryfirstbook} theory of polymer melts.  On the
cubic lattice, and in accord with simulations\cite{Enumeration},
the number of compact conformations is $\sim \exp (s N)$, where $s
\approx 0.62 $.  From that point of view, the above mentioned
result of trivial knot probability fitting to $\sim \exp(-N/196)$
implies that topologically restricting polymer to have the
topology of a trivial knot only reduces entropy per segment by
about $1/196 \approx 0.005$, which is a relatively insignificant
amount compared to the entropy $s \approx 0.62$ itself.  By
contrast, to obtain an estimate from the other end, we can
consider so-called crumpled conformations, similar to Peano
curves.  On the lattice, in the $2^k \times 2^k \times 2^k$ cube,
they can be defined in the following way: the $2^k \times 2^k
\times 2^k$ cube can be viewed as 8 smaller cubes $2^{k-1} \times
2^{k-1} \times 2^{k-1}$ each, and each smaller-subcube can be
further divided in a similar way, etc, down to the smallest $2
\times 2 \times 2$ cubes. We define the trajectory to be crumpled
if it visits all the vertices within given subcube before entering
next subcube of the same level.  It is easy to prove that crumpled
conformations are trivial knots. The exact recurrence relation can
be written\cite{my_unpublished_result} for the total number of
crumpled conformations for the polymer of $N=8^k$ monomers, it
yields the number of conformations about $\sim \exp (s^{\prime}
N)$, with $s^{\prime} \approx 0.30$.  Thus, there remains a huge
room for speculation regarding the asymptotic value of trivial
compact knot probability at $N \to \infty$.  Strictly speaking, we
cannot claim it is $196$ or close, we can only claim it is not
larger than this quantity.

The inset in figure~\ref{fig1} shows the probabilities of some
non-trivial knots. The existence of a maximum is easily explained
qualitatively as follows. When $N$ is small, the loop might be too
short for a given knot, i.e. there is not enough "room" for the
knot to exist. It is clear in the lattice model that there is a
finite number of segments required to form any given knot, e.g.
$N=24$ for a trefoil\cite{3}. At the other extreme, when $N$ is
large, there are many other knots "competing" for formation. The
number of complex knots increases with $N$, yielding a decaying
probability to locate any given knot.

\begin{figure}[th]      %Fig~2
%\centerline{\psfig{file=ratio.eps,width=2.6in,angle=0}}
\centerline{\scalebox{1}{\includegraphics{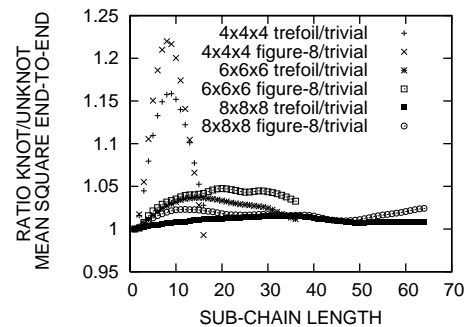}}}
\vspace*{8pt}
\caption{Mean square end-to-end of segments (sub-chains) of trefoil and figure-eight knotted conformations
 relative to that of trivially knotted conformations ($4\times4\times4,6\times6\times6,8\times8\times8$).
Pieces of knots are more extended compared to pieces of unknots.}\label{fig2}
\end{figure}

We also addressed the question of the spatial extent of segments of compact conformations. In particular, we were interested in
determining if the segments of knotted conformations are more stretched-out or more compact compared to segments of unknotted conformations.
(Note that even though the entire conformation is maximally compact, a connected piece of it need not be.)
To this end, we collected conformations of length $N$ containing a particular knot and measured the mean-square end-to-end distances for segments or sub-chains of length up to $N^{2/3}$. We found that segments of knotted conformations are consistently more spread-out on average compared to segments
of unknotted conformations. Figure~\ref{fig2} illustrates these results by plotting the ratios of mean-square end-to-end of segments
of trefoil and figure-eight knotted conformations to unknotted conformations.

\begin{figure}[th]      %Fig~3
%\centerline{\psfig{file=ratio2.eps,width=2.6in,angle=0}}
\centerline{\scalebox{1}{\includegraphics{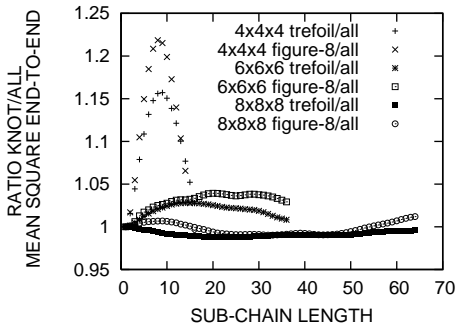}}}
\vspace*{8pt}
\caption{Mean square end-to-end of segments (sub-chains) of trefoil and figure-eight knotted conformations
 relative to that of the entire sample ("All"). Pieces of conformations that are over-knotted are more extended; Pieces of conformations
that are under-knotted are more compact.}\label{fig3}
\end{figure}

In figure~\ref{fig3}, instead of taking the ratio over trivial
knots, we took the ratio over all conformations in the sample
regardless of topology. Although figure~\ref{fig3} seems hardly
different from figure~\ref{fig2}, much of the points for the
$8\times8\times8$ trefoils and figure-eights in figure~\ref{fig3}
correspond to ratios less than unity. This means the trefoil and
figure-eight knotted conformations begin to be more compact than
typical conformations for size $8\times8\times8$, i.e. these knots
cross over from their over-knotted to their under-knotted regimes.
In fact, for $8\times8\times8$, the percentage of trivial, trefoil
and figure-eight  knots in a fairly generated sample of
conformations are 40\%, 20\% and 6.7\% respectively. The rest of
the more complex knots pull the average mean square end-to-end to
values larger compared to those of trivial, trefoil and
figure-eight knots. For smaller conformations of size
$4\times4\times4$ and $6\times6\times6$, the topology is dominated
by trivial knots.

\section{Testing knot localization hypothesis by renormalization}

In this section we attempt to address the issue of knot
localization for cicular chains using an idea inspired by field
theory and polymer physics called {\it renormalization} or {\it
decimation}\cite{4}. The idea is to group the $N$ segments of our
original circular chain into $N/g$ blocks or {\it blobs} of $g$
units each. A new circular chain is formed by connecting the
centers of mass of the blobs.

Our "renormalization" procedure works as follows. Starting from a
batch of chains of length $N$ with a given knot population
distribution, we renormalize each chain to obtain a new batch of
shorter chains of length $N_{1}=N/g$. We then compute the
probabilities to obtain various knots for the new batch of
renormalized or decimated chains and compare that to the knot
probabilities of the original batch. If the knots are localized,
we expect a renormalization step using a "small" value of $g$ to
obliterate or "wash out" any memory of the original knot state. In
other words, we expect a chain containing a localized knot and a
chain containing no knots (or a localized knot of another type) to
resemble each other after renormalization, with the value of $g$
giving an idea of the size of the knotted domain. For example, a
trefoil knot formed by six straight links (sticks) of equal length
will get totally unknotted after grouping consecutive units into
pairs ($g=2$) to yield a chain containing just three
(renormalized) units.

\begin{figure}[th]      %Fig~4
%\centerline{\psfig{file=trajsamplefinal.eps,width=2.6in,angle=0}}
\centerline{\scalebox{1}{\includegraphics{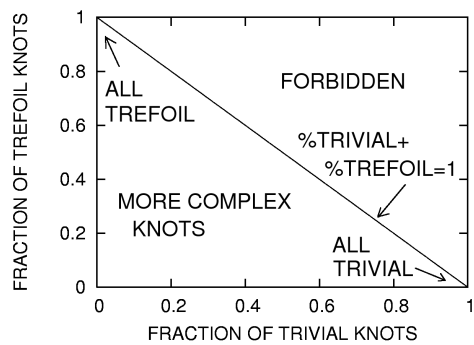}}}
\vspace*{8pt}
\caption{The space in which the trajectories are plotted. One can also consider it as a two dimensional slice
of the infinite dimensional space of knot probabilities.}\label{fig4}
\end{figure}

It is customary to present the results of a renormalization
procedure by constructing a trajectory. In our case, the axes of
the space in which the trajectory is plotted correspond to knot
probabilities. Since the number of distinct knots is infinite, the
dimension of this space is also infinite. To make the presentation
more manageable, we plot the trajectory in a plane with one axis
corresponding to the probability of trivial knots and the other
axis corresponding to the probability of trefoil knots
(Figure~\ref{fig4}). In this plane, the important regions are
marked. The point $(1,0)$ at the lower right hand corner of
figure~\ref{fig4} corresponds to all knots being trivial, while
the point $(0,1)$ at the upper left hand corner corresponds to
pure trefoil knots. The line connecting these two points going
diagonally corresponds to a sample containing either trivial and
trefoil but no other knots. The region above this line, marked
"FORBIDDEN", never gets visited by a trajectory by virtue of
probabilities always summing to $1$. The region below this line
corresponds to a sample containing more complex knots as well as
trivial and trefoil knots. By plotting the knot population in each
step (labeled by $g$) of the renormalization procedure, one gets a
picture of the evolution of knot complexity. After a sufficient
number of iterations or for large enough $g$, all trajectories
should terminate at the point $(1,0)$ corresponding to completely
unknotted chains.

\begin{figure}[th]      %Fig~5
%\centerline{\psfig{file=trajN384final.eps,width=2.6in,angle=0}}
\centerline{\scalebox{1}{\includegraphics{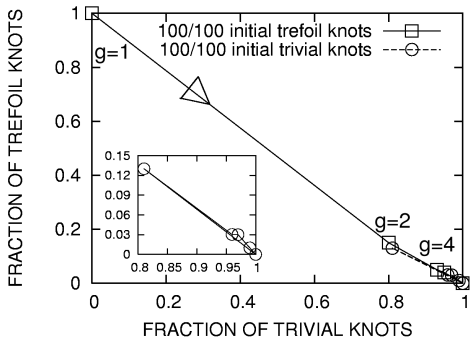}}}
\vspace*{8pt}
\caption{Knot renormalization trajectories for initial non-compact loops of length $N=384$.
Inset shows the trajectory only for initial unknotted loops.
(Initial trefoil knots - squares, initial trivial knots - circles).}\label{fig5}
\end{figure}

We first tested this procedure by examining non-compact circular loops of length $N=384$, using an algorithm\cite{5}
also described in a preceding article. We generated a batch of $100$ loops with trefoil knots and another batch of $100$ unknotted loops
or trivial knots. The loops in each batch were renormalized into blobs of size $g=2,4,8,16,32,64$ units and the fraction of trefoils and
trivial knots were computed for each $g$. When the probabilities for these knots are plotted against each other, one obtains
figure~\ref{fig5}.

In figure~\ref{fig5}, the trajectory for the $100$ trefoils (squares) starts at the top of the vertical axis while the
trajectory for the $100$ trivial knots (circles) starts at the extreme right along the horizontal axis. The trajectory of
the initial trefoils takes an almost straight path down towards the unknotted region (indicated by the arrow), implying that
the tendency to unknot is overwhelming (although a few non-trivial and non-trefoil knots were produced). The trajectory
for the initial trivial knots first takes a short step away from the starting point (circle, $g=2$), meaning some
knots predominantly trefoil are produced, then heads back towards the starting point.

The two trajectories first meet at $g=2$, when the chains become
roughly indistinguishable topologically. The fraction of trivial
knots at this point is $80\%$, which is significantly larger than
the probability of getting a trivial knot from a random sample of
loops with number of segments given by $N_{1}=N/g=384/2=192$. For
$N_0=241$, the empirical unknotting probability\cite{5} is given
by $w_{\rm triv}=\exp(-N_{1}/N_0)=45\%$. These results seem to be
consistent with the picture of a localized trefoil knot, using up
just a few segments (6 or 7) of the entire chain.

\begin{figure}[th]      %Fig~6
%\centerline{\psfig{file=traj8final.eps,width=2.6in,angle=0}}
\centerline{\scalebox{1}{\includegraphics{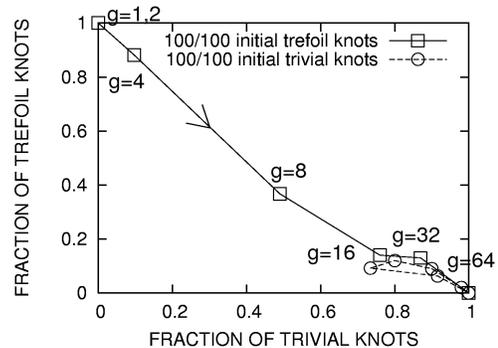}}}
\vspace*{8pt}
\caption{Knot renormalization trajectories for initial $8\times8\times8$ compact conformations.
(Initial trefoil knots - squares, initial trivial knots - circles).}\label{fig6}
\end{figure}

\begin{figure}[th]      %Fig~7
%\centerline{\psfig{file=traj12final.eps,width=2.6in,angle=0}}
\centerline{\scalebox{1}{\includegraphics{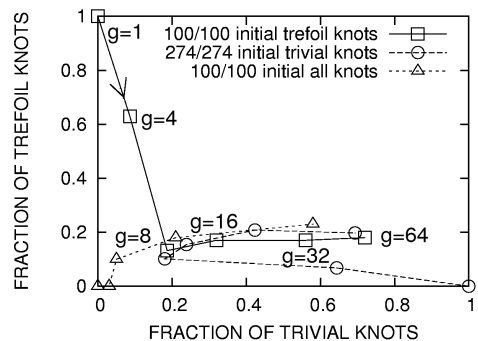}}}
\vspace*{8pt} \caption{Knot renormalization trajectories for
initial $12\times12\times12$ compact conformations. (Initial
trefoil knots - squares, initial trivial knots - circles, initial
"all" knots - triangles).}\label{fig7}
\end{figure}

The renormalization procedure outlined above was also applied to compact conformations on a
cubic lattice of dimensions $8\times8\times8$ and $12\times12\times12$ (figures~\ref{fig6} and~\ref{fig7}).
Figure~\ref{fig6} illustrates the trajectories for initial $8\times8\times8$ compact conformations of
$100$ trefoils and $100$ trivial knots. The trajectory of initial trefoils takes a downward
path slightly skewed towards the origin. The trajectory of initial trivial knots makes a short excursion
towards the origin then turns around and goes back to its starting point. The two trajectories seem to
meet at about $g=16$. The interpretation of a localized or delocalized knot is not straightforward, since
on a cubic lattice, the minimum number of segments needed to form a trefoil is $24$ instead of $6$.

The trajectories deviate significantly from the diagonal, manifesting the fact that many more complex
knots are formed as a result of the renormalization procedure. This result is not surprising, as the distance between renormalized segments
actually become smaller than the lattice constant of the original cubic lattice. In fact, renormalized segments may
even overlap, after say $g=4$, due to congruences arising from the integral coordinate positions of the
original chain units.

Figure~\ref{fig7} illustrates the trajectories for initial
$12\times12\times12$ compact conformations with $g=4,8,16,32,64$.
Aside from batches of 100 trefoils and 274 trivial knots, we also
added a batch of 100 conformations (triangles) regardless of
topology, representing typical compact conformations for this
size. In this "all" knots batch, the probability for either a
trivial knot or trefoil knot is about $2$ in $500$ (its starting
point is thus located at the origin). In this regime, trivial
knots and trefoil knots are overwhelmingly under-knotted. It can
be seen from the trajectory of "all" knots that the memory of the
initial knot state does persist longer the more complex the
initial knotted state. This result is also complemented by the
behaviour of the trajectories of initial trefoils and trivial
knots, which come close to the origin at $g=8$.

\section{Conclusion}

To conclude, we have examined the simple lattice model of compact
polymer loops.  To generate such loops computationally, we have
employed what we believe is the least biased algorithm currently
available\cite{1}.  We first confirmed that the probability to
realize a trivial knot is dramatically suppressed in compact loops
compared to their volume-unrestricted counterparts.  We sampled
compact loops of up to $N=14^3=2744$ segments, and our data fit $\sim
\exp(-N/N_0)$ with $N_0 \approx 196$.  We emphasize that it
remains unclear whether we have already reached the true $N \to
\infty$ asymptotics, or in fact $N_0$ is smaller than our observed
value of $196$.  When we looked at the end-to-end distance of the
small pieces of polymer, with some $g$ segments, $g \ll N^{2/3}$,
we found that their end-to-end distances are smaller when the loop
as a whole is a trivial knot, or, in general, when it is
under-knotted, compared to the average over all loops or compared
to complex knots or over-knotted loops.  In
principle, this allows for the possibility to make at least a probabilistic
judgment of the global loop topology by observing its local fractal
geometry.  This is also consistent, at least qualitatively, with
our findings in the preceding paper\cite{preceding} that
confinement entropy grows with decreasing size much sharper for
the topologically constrained loops such as trivial knots compared
at the topology-blind average over all loops.  We presented also
first attempts to address knot de-localization in the collapsed
polymers by looking at the renormalization trajectories in the
space of knots frequencies.  We think that our results open
prospects for deeper understanding of the interplay between knot
topology and loop compaction, and that finding a more solid
foundation for these studies is an acute challenge.

\section*{Acknowledgments}

We thank P. Pieranski for his use of SONO algorithm to confirm
trivial knot status of some of our biggest conformations, and also
for producing the image presented in Figure \ref{fig8}.  This work
was supported in part by the MRSEC Program of the National Science
Foundation under Award Number DMR-0212302.

\end{document}